\documentclass[aps,prb,reprint,twocolumn,superscriptaddress]{revtex4-2}
\usepackage{graphicx}
\usepackage{amsmath}
\usepackage{amsfonts}
\usepackage{color}
\usepackage{amssymb}%
\usepackage{siunitx}
\usepackage{blindtext}

\usepackage{xspace}
\usepackage{siunitx} 
\usepackage{miller}

\usepackage[version=3]{mhchem}
\usepackage{natbib}
\usepackage[]{hyperref}

\usepackage{ulem}
\usepackage{bm}

\renewcommand{\vec}[1]{{\boldsymbol #1}}

\newcommand{\mnwo}{MnWO$_4$ }
\newcommand{\mnwokomma}{MnWO$_4$}

\newcommand{\nh}{$\left(\text{NH}_4\right)_2\left[\text{FeCl}_5\left(\text{H}_2\text{O}\right)\right]$ }

\newcommand{\nafegeo}{NaFeGe$_2$O$_6$ }

\newcommand{\nafe}{NaFe$(\text{WO}_4)_2$ }
\newcommand{\nafekomma}{NaFe$(\text{WO}_4)_2$}
\newcommand{\life}{LiFe(WO$_4$)$_2$ }
\newcommand{\lifekomma}{LiFe$(\text{WO}_4)_2$}
\newcommand{\afe}{$A$Fe$(\text{WO}_4)_2$ }

\newcommand{\dmi}{Dzyaloshinskii-Moriya interaction }

\begin{document}


\title{Single-crystal investigations on the multiferroic material \life}

\author{S. Biesenkamp}\email[e-mail: ]{biesenkamp@ph2.uni-koeln.de}
\affiliation{$I\hspace{-.1em}I$. Physikalisches Institut,
Universit\"at zu K\"oln, Z\"ulpicher Stra\ss e 77, D-50937 K\"oln,
Germany}

\author{D. Gorkov}
\affiliation{$I\hspace{-.1em}I$. Physikalisches Institut,
Universit\"at zu K\"oln, Z\"ulpicher Stra\ss e 77, D-50937 K\"oln,
Germany}
\affiliation{FRM-II Munich,
Germany}

\author{D. Br\"uning}
\affiliation{$I\hspace{-.1em}I$. Physikalisches Institut,
Universit\"at zu K\"oln, Z\"ulpicher Stra\ss e 77, D-50937 K\"oln,
Germany}

\author{A. Bertin}
\affiliation{$I\hspace{-.1em}I$. Physikalisches Institut,
Universit\"at zu K\"oln, Z\"ulpicher Stra\ss e 77, D-50937 K\"oln,
Germany}

\author{T. Fr\"ohlich}
\affiliation{$I\hspace{-.1em}I$. Physikalisches Institut,
Universit\"at zu K\"oln, Z\"ulpicher Stra\ss e 77, D-50937 K\"oln,
Germany}

\author{X. Fabr\`{e}ges}
\affiliation{Laboratoire L\'eon Brillouin, C.E.A./C.N.R.S., F-91191 Gif-sur-Yvette CEDEX, France}

\author{A. Gukasov}
\affiliation{Laboratoire L\'eon Brillouin, C.E.A./C.N.R.S., F-91191 Gif-sur-Yvette CEDEX, France}

\author{M. Meven}
\affiliation{RWTH Aachen University, Institut f\"ur Kristallographie, 52056 Aachen, Germany}
\affiliation{J\"ulich Centre for Neutron Science JCNS at Heinz Maier-Leibnitz Zentrum (MLZ)}

\author{P. Becker}
\affiliation{Abteilung  Kristallographie,  Institut  f\"ur  Geologie  und  Mineralogie, Universit\"at zu K\"oln, Z\"ulpicher Straße  49b,  50674  K\"oln,  Germany}

\author{L. Bohat\'{y}}
\affiliation{Abteilung  Kristallographie,  Institut  f\"ur  Geologie  und  Mineralogie, Universit\"at zu K\"oln, Z\"ulpicher Straße  49b,  50674  K\"oln,  Germany}

\author{T. Lorenz}
\affiliation{$I\hspace{-.1em}I$. Physikalisches Institut,
Universit\"at zu K\"oln, Z\"ulpicher Stra\ss e 77, D-50937 K\"oln,
Germany}

\author{M. Braden}\email[e-mail: ]{braden@ph2.uni-koeln.de}
\affiliation{$I\hspace{-.1em}I$. Physikalisches Institut,
Universit\"at zu K\"oln, Z\"ulpicher Stra\ss e 77, D-50937 K\"oln,
Germany}





\date{\today}

\begin{abstract}
The crystal and magnetic structure of multiferroic \life were investigated by temperature and magnetic-field dependent specific heat, susceptibility and neutron diffraction experiments on single crystals. Considering only the two nearest-neighbour magnetic interactions, the system forms a $J_1$, $J_2$ magnetic chain but more extended interactions are sizeable. Two different magnetic phases exhibiting long-range incommensurate order evolve at $T_{\text{N}1}\approx\SI{22.2}{\kelvin}$ and $T_{\text{N}2}\approx\SI{19}{\kelvin}$. First, a spin-density wave develops with moments lying in the $ac$ plane. In its multiferroic phase below $T_{\text{N}2}$, \life exhibits a spiral arrangement with an additional spin-component along $b$. Therefore, the inverse Dzyaloshinskii-Moriya mechanism fully explains the multiferroic behavior in this material. A partially unbalanced multiferroic domain distribution was observed even in the absence of an applied electric field. For both phases only a slight temperature dependence of the incommensurability was observed and there is no commensurate phase emerging at low temperature or at finite magnetic fields up to $\SI{6}{\tesla}$. \life  thus exhibits a simple phase diagram with the typical sequence of transitions for a type-II multiferroic material.
\end{abstract}


\maketitle


\section{\label{sec:level1}Introduction}
The demand of memory devices with larger storage capacity and less power consumption pushed the research on multiferroic materials featuring the coupling of magnetic ordering and ferroelectric polarization in the same phase (so called type-II multiferroics)\cite{Scott2007a,Fiebig_2005,Spaldin2019}. During the past decades a variety of multiferroic systems were discovered and several different microscopic mechanisms were identified to induce multiferroicity in respective materials \cite{Fiebig2016,Khomskii2009}. 
Beside symmetric exchange striction, the inverse \dmi (DMI) is a frequent mechanism that drives the multiferroic behavior in many type-II multiferroics \cite{Dzyaloshinsky1958,Moriya1956,Mostovoy2006,Kimura2007}. With this mechanism, a spiral spin canting of neighboring spins induces a shift of nonmagnetic ligand ions and thus a ferroelectric polarization that can be controlled by external electric and magnetic fields. The spiral handedness (or precisely the vector chirality) determines the sign of the ferroelectric polarization.

Many multiferroic materials exhibit additional phase transitions at low temperature either due to other magnetic constituents or due to locking into a commensurate phase. Anharmonic modulations of the spiral structure also cause sizable magnetoelastic coupling, which is different to the multiferroic one and which can lead to anomalous relaxation behavior of multiferroic domains\cite{Baum2014,Biesenkamp2020}.  Beside the demand for higher transition temperatures and larger ferroelectric polarization it is also necessary to find multiferroic materials exhibiting simpler phase diagrams in order to analyze and describe multiferroic domain dynamics.

%

\begin{figure}
 \includegraphics[width=\columnwidth]{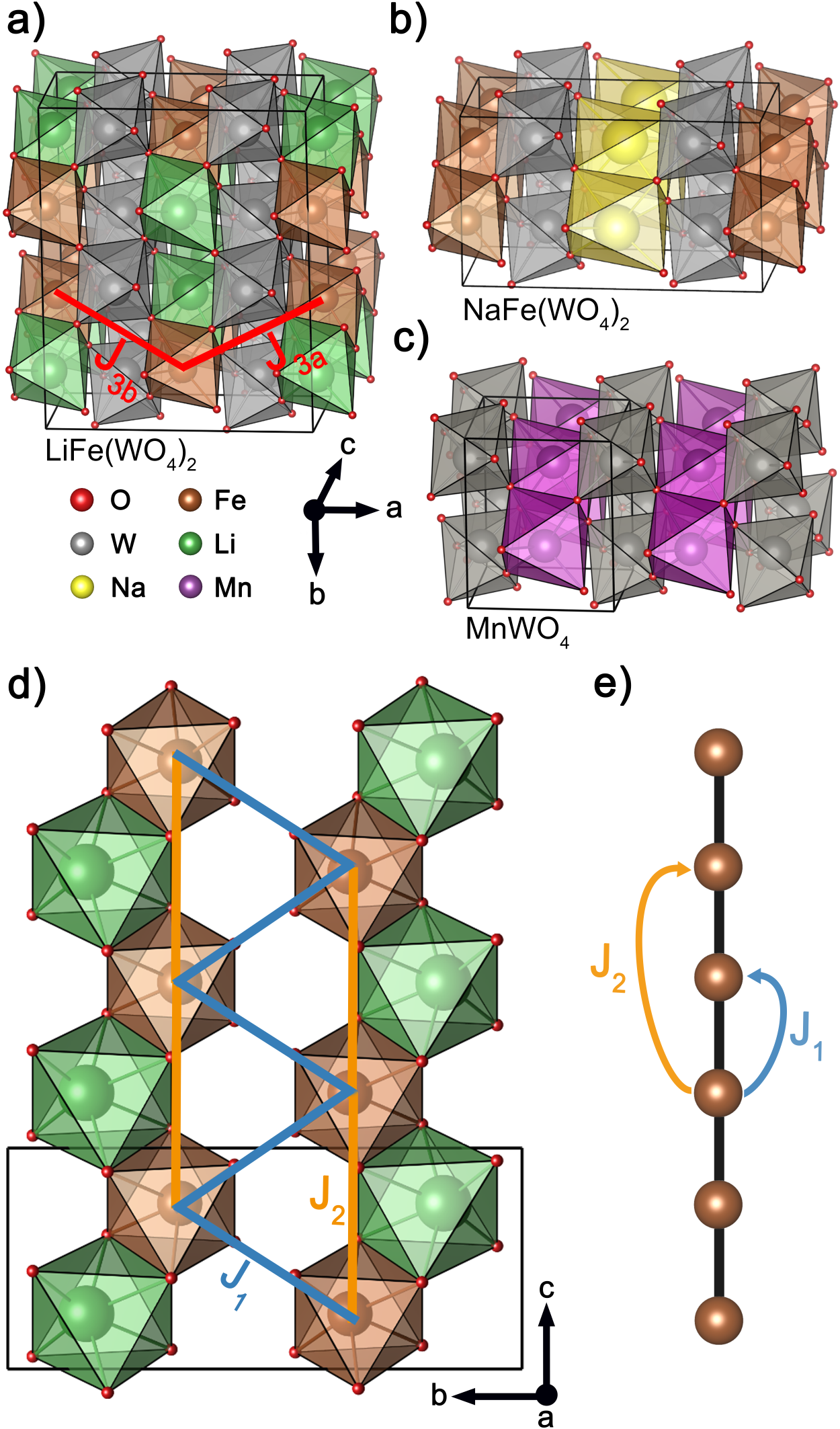}
  \caption{\label{fig:structure}
  Crystal structure and magnetic interaction paths in \lifekomma{.}  Figure a) displays the nuclear structure of \lifekomma{,} for which the structural parameters were adopted from the refinement results  based on X-ray diffraction data at $\SI{290}{\kelvin}$ (see table \ref{tab:table_nuc_refinement}). The exchange couplings $J_{3a}$ and $J_{3b}$ are marked in red. For comparison the nuclear structure of related \nafe and \mnwo is shown in b) and c). In d) the magnetic zigzag chains along $c$ are visualized together with the respective exchange couplings $J_1$ and $J_2$. For the sake of simplicity tungsten ions are hidden in d). Figure e) visualizes the configuration of nearest-neighbor interaction $J_1$ and next-nearest neighbor interaction $J_2$ along the magnetic Fe$^{3+}$ chain. For the crystal structure visualization the software \textsc{VESTA3} \cite{Momma2011} was utilized.}
  \label{disp}
 \end{figure}

Recently, Liu \textit{et. al.} reported a type-II multiferroic phase in \life\cite{Liu2017}, which is beside \mnwo\cite{Taniguchi2006,Heyer2006} only the second multiferroic material in the family of tungstates and which  exhibits higher transition temperatures compared to related compounds \cite{Anders1975,Kimura2003,Lawes2005,Ackermann2013,Ackermann_2015,Jodlauk_2007,Holbein2016,Taniguchi2006,Heyer2006}. In \mnwokomma{,} the magnetic ions occupy zigzag chains that are lying within the $bc$ plane \cite{Lautenschlager1993}, see Fig. 1. These chains propagate along $c$ direction and are separated along $a$ by tungsten layers \cite{Lautenschlager1993}.
In the case of the double tungstate \nafekomma{,} the magnetic ion is substituted by Na in every second chain stacked along the $a$ axis of the \mnwo structure type.
Therefore, the $a$ lattice constant as well as the distance between two magnetic zigzag chains along $a$ are doubled lowering the dimensionality of magnetic interaction (see Fig. \ref{fig:structure}) \cite{Holbein2016}. Distinctive for \life is that the zigzag chains are not entirely occupied by magnetic ions but alternatingly by magnetic Fe$^{3+}$ and by nonmagnetic Li$^{1+}$ ions (see Fig. \ref{fig:structure})\cite{Liu2017,Flem1969,Klevtsov1970}. Similar to \mnwo every second layer along $a$ is magnetic in \lifekomma{.}

Two successive magnetic anomalies have been reported for \life ~ at $T_{\text{N}1}\approx\SI{22.6}{\kelvin}$ and $T_{\text{N}2}\approx\SI{19.7}{\kelvin}$ and the first one was associated with the onset of short-range ordering \cite{Liu2017}. At the lower transition an emerging ferroelectric polarization of about $\SI{15}{\micro\coulomb\per\square\metre}$ was observed but the corresponding measurements were performed on polycrystalline samples thus preventing the determination of the polar axis \cite{Liu2017}. Density-functional theory (DFT) calculations suggest a ferroelectric polarization along the $b$ direction, whose magnitude is comparable to the one observed in \mnwo \cite{Liu2017,Taniguchi2006,Arkenbout2006}. Neutron powder diffraction (NPD) revealed an incommensurate spiral ordering of magnetic moments below $T_{\text{N}2}$ and thus suggests the inverse \dmi DMI as the underlying mechanism for multiferroicity in  \life \cite{Liu2017}. No transition to a commensurate phase was observed and specific heat measurements reported only a slight magnetic-field dependence of both transitions indicating strong antiferromagnetic coupling as well as a simple phase diagram of \life \cite{Liu2017}.

However, NPD experiments are not sufficient to fully characterize complex magnetic structures in form of incommensurate spiral arrangements. Furthermore, it is unusual for spiral type-II multiferroics that short range magnetic ordering directly turns into a long-range spiral arrangement of magnetic moments without an intermediate phase \cite{Mostovoy2006,Toledano2009,Toledano2010}. For multiferroic MnWO$_4$, various magnetic and nonmagnetic chemical substitutions have been studied and the intermediate antiferromagnetic paraeletric phase is not suppressed for Fe\cite{Ye2008}, Co\cite{Liang2012}, Ni\cite{Poudel2015}, Cu\cite{Kumar2015}, Zn\cite{Meddar2009}, Mg\cite{Meddar2009}, Mo\cite{Meddar2012}, In\cite{Gattermann2016} and Ir\cite{Wang2019} substitution.
For \life a thorough investigation of the first magnetic transition at $T_{\text{N}1}$ and confirmation of the spiral character below $T_{\text{N}2}$ are thus needed requiring single-crystal investigations.
Here we report on single-crystal investigations concerning the magnetization, the specific heat at magnetic field, and the nuclear and magnetic structure utilizing different experimental techniques. After the introduction to experimental methods we will first discuss specific-heat data and the temperature-dependent refinements of the nuclear structure before we subsequently discuss both observed magnetic phases, which have been characterized by susceptibility measurements, structural refinements and by neutron polarization analysis.

  \section{\label{sec:exp_method}Experimental Methods}

Single crystals of about $\SI{3}{\milli\metre}\times\SI{3}{\milli\metre}\times\SI{0.5}{\milli\metre}$ in size were grown from a lithium poly-tungstate melt solution. The samples exhibit a finite conductivity even at low temperature, which prevents the application of electric fields and the direct investigation of the ferroelectric polarization. The measurements, which are discussed below in the main section, were performed on the same prepared single crystal SI. The investigations on a second sample SII exhibiting reduced transition temperatures are reported in the appendix \ref{sec:appendix}.

The characterization of the low-temperature phase transitions by susceptibility and by specific-heat measurements were done on a commercial superconducting quantum interference device (SQUID) magnetometer and on a commercial calorimeter (PPMS, Quantum Design) by using the thermal relaxation-time method, respectively. For the nuclear and magnetic structure determination, diffraction experiments with X-ray and neutron radiation were executed. The respective single-crystal X-ray data collection for a refinement of the nuclear structure was carried out on a Bruker AXS Kappa APEX II four-circle X-ray diffractometer with  MoK$_\alpha$ radiation ($\lambda=\SI{0.71}{\angstrom}$). An Oxford N-HeliX cryosystem was additionally deployed for low-temperature measurements. A collection of magnetic reflections as well as temperature and magnetic-field dependent $Q$ space mappings of magnetic reflections were undertaken on the diffractometer 6T2, which is located at the Laboratoire L\'eon Brillouin (LLB). The instrument was equipped with a vertical cryomagnet $(\mu_0H\leq\SI{6}{\tesla})$, a lifting counter detector and a vertically focusing  pyrolytic graphite monochromator yielding a wavelength $\lambda=\SI{2.35}{\angstrom}$. A neutron diffraction experiment utilizing a polarized neutron beam was executed on the cold neutron three axes spectrometer KOMPASS, which is located at the Heinz Maier-Leibnitz Zentrum (MLZ). Serial polarizing V-shaped multichannel cavities provided an incoming polarized neutron beam and a highly oriented pyrolytic graphite (HOPG(002)) monochromator was used to select neutrons with $\lambda=\SI{4}{\angstrom}$. A Helmholtz-Coil setup for defining the guide-field direction at the sample position was deployed together with a secondary V-cavity polarizer in front of the detector thus enabling longitudinal polarization analysis.

 \section{\label{sec:level1}Results and discussion}
 \subsection{\label{sec:nuclear_structure}Nuclear structure}
The nuclear structure was determined from X-ray and neutron diffraction data as a function of temperature.  All atomic positions and isotropic displacement factors were refined by using the software \textsc{Jana2006} \cite{Petricek2014} and the corresponding results are summarized in table \ref{tab:table_nuc_refinement}.

The crystal structure was first discussed in Ref. \onlinecite{Flem1969} and it was reported that \life crystallizes in the monoclinic space group $C2/c$ \cite{Klevtsov1970,Anders1975}, which is confirmed by our measurements. No superstructure reflections were observed as a function of temperature indicating no violation of the $C$ centering or a structural transition. The system consists of zigzag chains that are propagating along $c$ direction (see Fig. \ref{fig:structure}). In contrast to \nafe the nonmagnetic monovalent alkali-metal ion $A$ in \afe and the magnetic Fe$^{3+}$ ions are not solely occupying the zigzag chains in \life but alternating within the same chain. With respect to \mnwokomma{,} the cell is not only doubled along $a$ direction as it is the case in \nafe but also along the $b$ direction due to a phase shift of the Li$^{1+}$ and Fe$^{3+}$ ordering. In \nafe the [WO$_6$] octahedra layer separates nonmagnetic [NaO$_6$] and magnetic [FeO$_6$] layers, whereas in \life the layers of [WO$_6$] units alternate with layers that contain both, Li$^{1+}$ and Fe$^{3+}$ ions.

By lowering the temperature to $T=\SI{100}{\kelvin}$ the respective inter- and intra-layer distances shrink by $\approx\SI{0.3}{\percent}$, whereas a refinement at $T=\SI{38}{\kelvin}$ reveals no further significant length changes.
We do not find any evidence for a structural phase transition and the low-temperature atomic displacement parameters are not enhanced.

\begin{table}
\caption{\label{tab:table_nuc_refinement}This table contains the temperature dependent refinements of the nuclear structure based on single crystal X-ray data together with resulting reliability factors for structure factors. The temperature dependent lattice constants have been obtained by refining the orientation matrix considering the complete set of recorded X-ray data at the respective temperature.  All refinements were  done by using the software package \textsc{Jana2006} \cite{Petricek2014} and by assuming space group $C2/c$.}
\begin{ruledtabular}
\begin{tabular}{m{0.9cm}ccccc}

 &&x  &y  & z &U$_\text{iso}$\\ \hline

$\SI{290}{\kelvin}$&W&0.247505(9)&0.091348(9)&0.246221(18)&0.00299(3)\\
	  &Fe&0.0  &0.33485(5)  &0.25  &0.00447(8)\\
      &Li&0.5  &0.3421(7)   &0.25  &0.008(2)\\
      &O1&0.3634(2)  &0.05909(18)  &0.9240(4)  &0.0052(3)\\
      &O2&0.3801(2)  &0.18198(19)  &0.4114(4)  &0.0056(3)\\
      &O3&0.3552(2)  &0.54862(18)  &0.9452(4)  &0.0052(3)\\
      &O4&0.3769(2)  &0.69430(19)  &0.3928(4)  &0.0059(3)\\
 &\multicolumn{5}{m{7cm}}{\centering Recorded reflections: 70005, Independent: 2816 $a$=9.2894(5), $b$=11.4142(6), $c$=4.9026(3), $\beta$=90.574(2), R(obs)=2.71, wR(obs)=3.08, R(all)=3.64, wR(all)=3.23
}\\
  &&&&&\\
$\SI{100}{\kelvin}$&W&0.24709(3)&0.09137(2)&0.24661(6)&0.00193(9)\\
	  &Fe&0.0  &0.33466(13)  &0.25  &0.0019(2)\\
      &Li&0.5  &0.344(3)   &0.25  &0.032(7)\\
      &O1&0.3635(6)  &0.0586(5)  &0.9252(11)  &0.0032(8)\\
      &O2&0.3800(7)  &0.1830(5)  &0.4121(12)  &0.0063(10)\\
      &O3&0.3557(6)  &0.5485(5)  &0.9446(11)  &0.0047(9)\\
      &O4&0.3781(6)  &0.6946(5)  &0.3936(11)  &0.0037(9)\\
 &\multicolumn{5}{m{7cm}}{\centering Recorded reflections: 14973, Independent: 1595 $a$=9.252(4), $b$=11.383(4), $c$=4.8897(18), $\beta$=90.44(2), R(obs)=4.00, wR(obs)=3.67, R(all)=6.79, wR(all)=4.29
}\\
  &&&&&\\
$\SI{38}{\kelvin}$&W&0.24703(3)&0.09137(2)&0.24675(6)&0.00226(8)\\
	  &Fe&0.0  &0.33474(13)  &0.25  &0.0025(2)\\
      &Li&0.5  &0.346(2)   &0.25  &0.022(5)\\
      &O1&0.3638(6)  &0.0598(5)  &0.9245(10)  &0.0050(8)\\
      &O2&0.3801(6)  &0.1822(5)  &0.4115(10)  &0.0051(8)\\
      &O3&0.3562(5)  &0.5497(5)  &0.9437(10)  &0.0045(8)\\
      &O4&0.3775(5)  &0.6951(5)  &0.3933(10)  &0.0046(8)\\
 &\multicolumn{5}{m{7cm}}{\centering Recorded reflections: 17924, Independent: 1681 $a$=9.2648(11), $b$=11.3858(13), $c$=4.8918(6), $\beta$=90.400(7), R(obs)=3.57, wR(obs)=3.45, R(all)=6.30, wR(all)=4.04
}\\
\end{tabular}
\end{ruledtabular}
\end{table}

  \subsection{\label{sec:specific_heat}Specific heat}
\begin{figure}
 \includegraphics[width=\columnwidth]{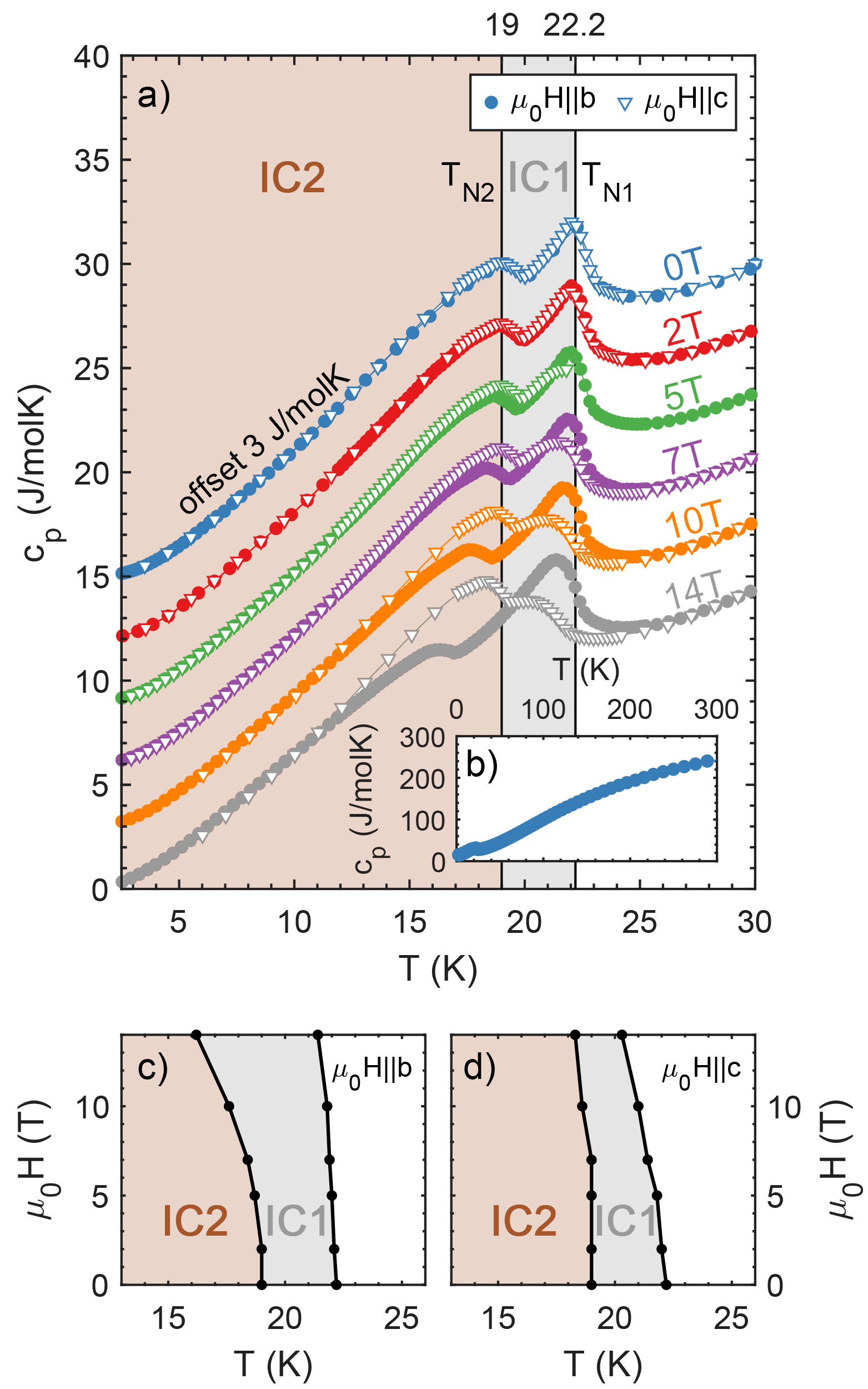}
  \caption{\label{fig_cp}In a), the specific heat of \life in magnetic fields along $b$ and $c$ is shown. Different fields are offset by 3\,J/molK. Both phases IC1 and IC2 are marked by colored regions and the respective transition temperatures $T_{\text{N}1}$ and $T_{\text{N}2}$ refer to the zero-field behavior. The zero-field data for temperatures up to room temperature are shown in the inset b). Both panels c) and d) display the obtained phase boundaries of the IC1 and IC2 phase for different field directions.}
 \end{figure}

Fig. \ref{fig_cp} displays the temperature dependence of the specific heat for different magnetic fields  applied either along $b$ or along $c$. The $c_p$ anomalies located at $T_\text{N1}\approx$ 22.2\,K and $T_\text{N2}\approx$ 19.0\,K agree with those from Liu \textit{et. al.} \cite{Liu2017}. As it can be seen in the inset b) of Fig. \ref{fig_cp}, no further anomalies were observed for higher temperatures up to room temperature. In magnetic fields up to 14\,T along $b$, $T_\text{N1}$ is only weakly suppressed, while $T_{\text{N}2}$ is significantly reduced to $\SI{16.2}{\kelvin}$, see Fig. 2 c). This indicates that ordering of $b$ components of the magnetic moments
takes not place at the upper but only at the lower transition.
In contrast a magnetic field along $c$ causes a more pronounced decrease of $T_\text{N1}$, but a weaker decrease of $T_\text{N2}$ see Fig. 2 d) indicating
ordering of $c$ components at the upper transition.
This anisotropic behavior clearly documents different magnetic ordering phenomena to occur at $T_\text{N1}$ and $T_\text{N2}$ and thus the presence of two different phases IC1 and IC2, which are studied in the following.
 \subsection{\label{sec:Propagation_vector}Susceptibility}

\begin{figure}
 \includegraphics[width=\columnwidth]{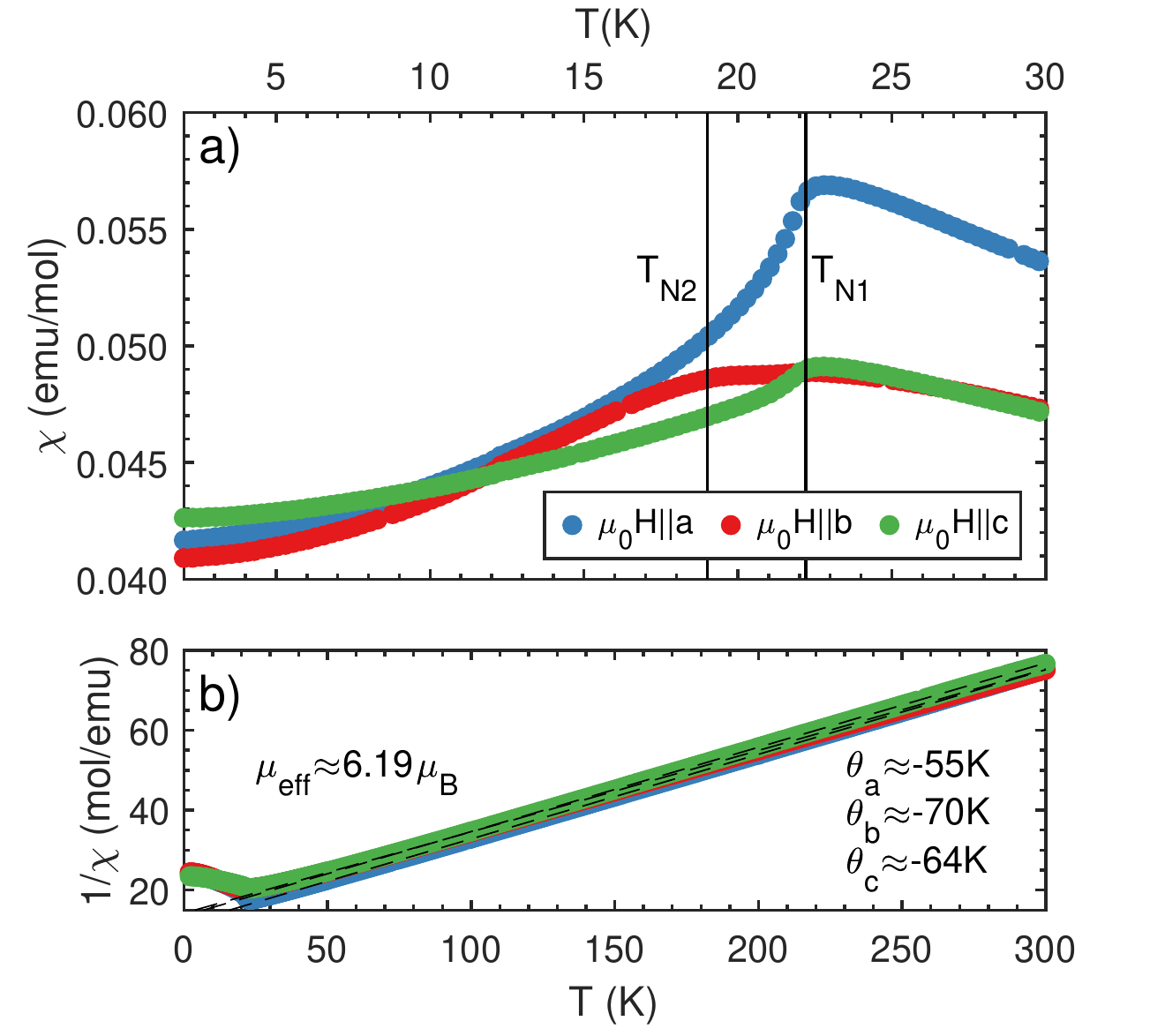}
  \caption{\label{fig:susceptibility}Fig. a) displays the temperature dependence of the susceptibility $\chi$ for all crystallographic directions around the phase transitions $T_{\text{N}1}$ and $T_{\text{N}2}$. The susceptibility was measured after zero-field cooling, while heating and applying a constant field of $\mu_0H=\SI{0.1}{\tesla}$ to the sample. Curie-Weiss fits for $1/\chi$ are shown by dashed black lines in figure b) together with the resulting Weiss temperatures $\theta$ and the averaged effective moment $\mu_{\text{eff}}$.}
  \label{disp}
 \end{figure}

The susceptibility was measured for magnetic fields applied along all crystallographic directions. As the crystal structure of \life is monoclinic, the $a$ direction is not equivalent to $a^*$, but because the monoclinic angle is close to $\SI{90}{\degree}$, the effect on the respective orientation is negligible. The same holds for $c$ and $c^*$. A detailed view of the low-temperature region is displayed in Fig. \ref{fig:susceptibility}a). At $T_{\text{N}1}\approx\SI{22.2}{\kelvin}$ the susceptibility exhibits a kink indicating the onset of long-range magnetic ordering. The susceptibility for fields applied along the $a$  and $c$ directions starts to decrease below $T_{\text{N}1}$, whereas for a field applied along $b$ direction, the susceptibility remains at the same level and decreases only below the second magnetic transition at $T_{\text{N}2}\approx\SI{19}{\kelvin}$. Both transition temperatures agree with the respective temperatures of the specific-heat anomalies in Fig. \ref{fig_cp}, but $T_{\text{N}1}$ and $T_{\text{N}2}$ are reduced by about $\SI{0.6}{\kelvin}$ compared to the transitions temperatures reported by Liu \textit{et. al.} \cite{Liu2017}. The temperature-dependent susceptibility data suggests that magnetic moments are initially aligning within the $ac$ plane and that an additional $b$ component of the magnetic ordering is evolving below $T_{\text{N}2}$. This agrees with the different impact of magnetic fields along $b$ and $c$ on $T_{\text{N}1}$ and $T_{\text{N}2}$ (see Fig. \ref{fig_cp}). The inverse susceptibility of the high temperature region was fitted by the Curie-Weiss function $C^{-1}\left(T+\theta\right)$ yielding antiferromagnetic Weiss temperatures for different field directions in the range of $\theta\approx\SI{-70}{\kelvin}$ to $\SI{-55}{\kelvin}$ (see Fig. \ref{fig:susceptibility}b)).
 From this, the averaged effective magnetic moment $\mu_\text{eff}\approx 6.19\mu_\text{B}$ was calculated and its value agrees well with the expected spin-only moment $\mu_\text{eff}=2\sqrt{S\left(S+1\right)}=5.91\mu_\text{B}$ for Fe$^{3+}$ $\left(S=5/2, L=0\right)$. The frustration parameter $f=\left|\theta\right|/T_N\approx 2.84$ is comparable to that in \nafe $\left(f\approx 2\text{, see reference \onlinecite{Holbein2016}}\right)$ and signals a moderate frustration. For comparison, \mnwo exhibits a larger frustration parameter of about 6 (see reference \onlinecite{Taniguchi2006}).  In contrast to \nafe no broad maximum was observed in \life above the first magnetic transition thus indicating a more three-dimensional magnetic interaction. Compared to the low-dimensionality in \nafe this is not astonishing, as the interlayer distance along $a$ is significantly reduced, which enhances the magnetic interaction between layers and explains the higher antiferromagnetic transitions.

\begin{figure}
 \includegraphics[width=\columnwidth]{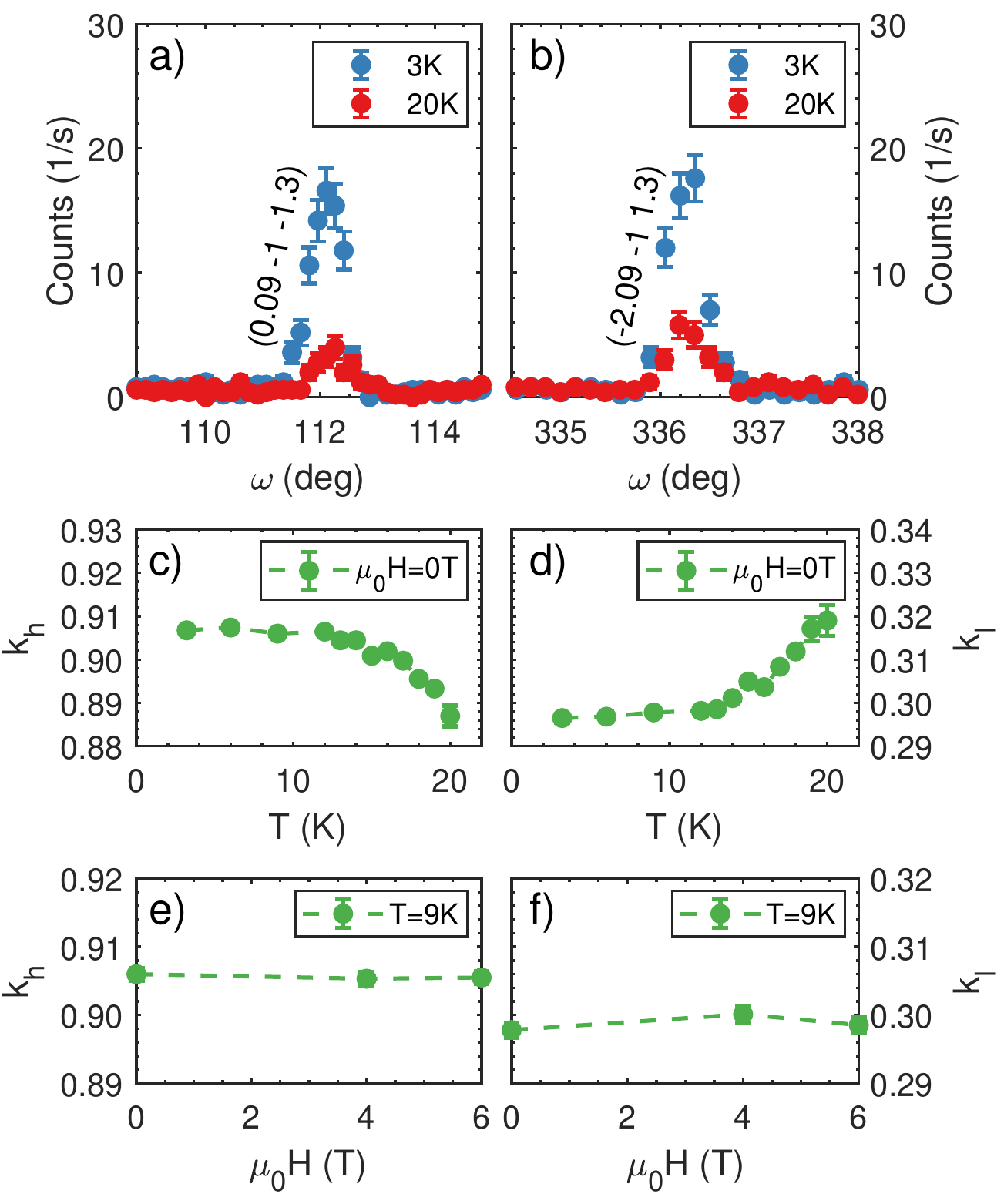}
  \caption{\label{fig:kvector}Fig. a) and b) display rocking scans over the magnetic reflections $\vec{Q}=\left(0.09\:\mathrm{-}1\:\mathrm{-}1.30\right)$ and $\vec{Q}=\left(-2.09\:\mathrm{-}1\:1.30\right)$ for $T=\SI{3}{\kelvin}$ and $T=\SI{20}{\kelvin}$ respectively. The temperature and magnetic field dependence of the incommensurate propagation vector $\vec{k_\text{inc}}=\left(k_h\:0\:k_l\right)$ is shown in c) - f). All respective values have been obtained by fitting the $\vec{Q}$-space mappings in the $h$-$l$ plane with a two-dimensional Gaussian function. The magnetic field was applied along $b$ direction.}
  \label{disp}
 \end{figure}

 \subsection{\label{sec:kvector}Incommensurate propagation vector}

For tracing the temperature and magnetic-field dependence of evolving magnetic reflections and that of the incommensurability in \lifekomma{} we performed experiments on the 6T2 diffractometer. A vertical cryomagnet ($\mu_0H\leq\SI{6}{\tesla}$) was utilized together with a lifting counter detector that allows for the collection of reflections with a  finite $k$ component of the scattering vector $\vec{Q}$. For the respective experiment, the sample was mounted in the $\left(1\:0\:0\right)/\left(0\:0\:1\right)$ scattering plane thus aligning the $b$ direction parallel to the magnetic-field.

Fig. \ref{fig:kvector} a) and b) display rocking scans over the magnetic reflections $\vec{Q}=\left(0.09\:\mathrm{-}1\:\mathrm{-}1.3\right)$ and $\vec{Q}=\left(\mathrm{-}2.09\:\mathrm{-}1\:1.3\right)$ for $T=\SI{3}{\kelvin}$ and $T=\SI{20}{\kelvin}$, respectively. The indexing of both reflections yields similar $h$  and $l$ components of the incommensurate propagation vector $\vec{k_{\text{inc}}}=\left(0.91\:0\:0.30\right)$ that qualitatively agree with reference \onlinecite{Liu2017},
who report an almost commensurate $l$=0.332 component. For both magnetic phases IC1 and IC2 a distinct peak is visible at the respective position in $Q$ space. From this it can be stated that long-range magnetic order also persists in the IC1 phase. To trace the incommensurability in the respective magnetic phases, $\vec{Q}$-space mappings within the $h$-$l$ plane were recorded around magnetic reflections as a function of temperature and magnetic field. Fig. \ref{fig:kvector} c) and d) show the temperature dependence of the resulting averaged $h$ and $l$ component of the incommensurate propagation vector for $\mu_0H=\SI{0}{\tesla}$ and it can be clearly seen that below $T_{\text{N}1}$ the components of $\vec{k_\text{inc}}$ vary with temperature but reach a constant value at low temperature. Incommensurate magnetic long-range order persists down to low temperature contrarily to \mnwokomma{,} which exhibits a first-order transition to a commensurate spin up-up-down-down (uudd) phase at low temperature \cite{Holbein2016,Lautenschlager1993}.

The magnetic-field dependence of the incommensurability at $T=\SI{9}{\kelvin}$ is presented in Fig. \ref{fig:kvector} e) and f). In contrast to related \nafekomma{,} no variation of $k_h$ and $k_l$ is visible as a function of magnetic field strength. The absence of field-dependent alteration of the incommensurability in \life further illustrates a well-defined antiferromagnetic coupling in accordance with the simplicity of its phase diagram. Just two incommensurate magnetic phases exist at low temperature and persist for finite magnetic fields along $b$ and $c$ direction. The determination of the appropriate magnetic models and the refinement of the respective magnetic structure will be discussed in the following sections.

\begin{table}
\caption{\label{tab:character_table}  Symmetry conditions for the transformation of the complex amplitudes $u$, $v$ and $w$ between both magnetic sites for the little group $G_\vec{k_\text{ic}}=\left\{1,c\right\}$. The conditions are defined by the irreducible representation $\Gamma_1$ and $\Gamma_2$ and the corresponding basis vectors $\Psi$ with  $a = e^{-i2\pi0.15}$. Equivalent results were reported in Ref. \onlinecite{Liu2017}}.\begin{ruledtabular}
\begin{tabular}{cccc}

 $\Gamma$&$\Psi$&($x$,$y$,$z$)&($x$,$\bar{y}$,$z+1/2$)\\ \hline

$\Gamma_1$&$\Psi_1$,$\Psi_2$,$\Psi_3$&($u$,$v$,$w$)&$a$($u$,$-v$,$w$)\\
$\Gamma_2$&$\Psi_4$,$\Psi_5$,$\Psi_6$&($u$,$v$,$w$)&$a$($-u$,$v$,$-w$)\\

\end{tabular}
\end{ruledtabular}
\end{table}

 \subsection{\label{sec:mag_structure_af1_af2}Magnetic structure of phase IC1 and IC2}
In \lifekomma, the propagation vector remains incommensurate down to the lowest temperature (see section \ref{sec:kvector}). $\vec{k_\text{inc}}$ and -$\vec{k_\text{inc}}$ are not equivalent for both magnetic phases and thus  two vectors belong to the star of $\vec{k}$. By considering space group $C2/c$ and the incommensurate propagation vector $\vec{k_\text{ic}}=\left(0.91\:0\:0.30\right)$, the little group $G_\vec{k_\text{ic}}=\left\{1,c\right\}$ can be deduced. The corresponding magnetic representations can be decomposed into two one-dimensional irreducible representations $\Gamma_\text{mag}=3\Gamma_1+3\Gamma_2$. Both magnetic Fe sites on the 4f Wyckoff position are linked via a $c$ glide-plane symmetry and thus belong to one orbit. Therefore, only three complex amplitudes $u$,$v$,$w$ are needed to describe the magnetic moment on each site. The phase difference between the two sites amounts $\phi_\vec{k}=2\pi\times 0.15$ and the symmetry restrictions referring to the transformation of magnetic moments between the respective sites $(x,y,z)$ and ($x$,$\bar{y}$,$z+1/2$) are displayed in table \ref{tab:character_table}.

\begin{figure}
 \includegraphics[width=\columnwidth]{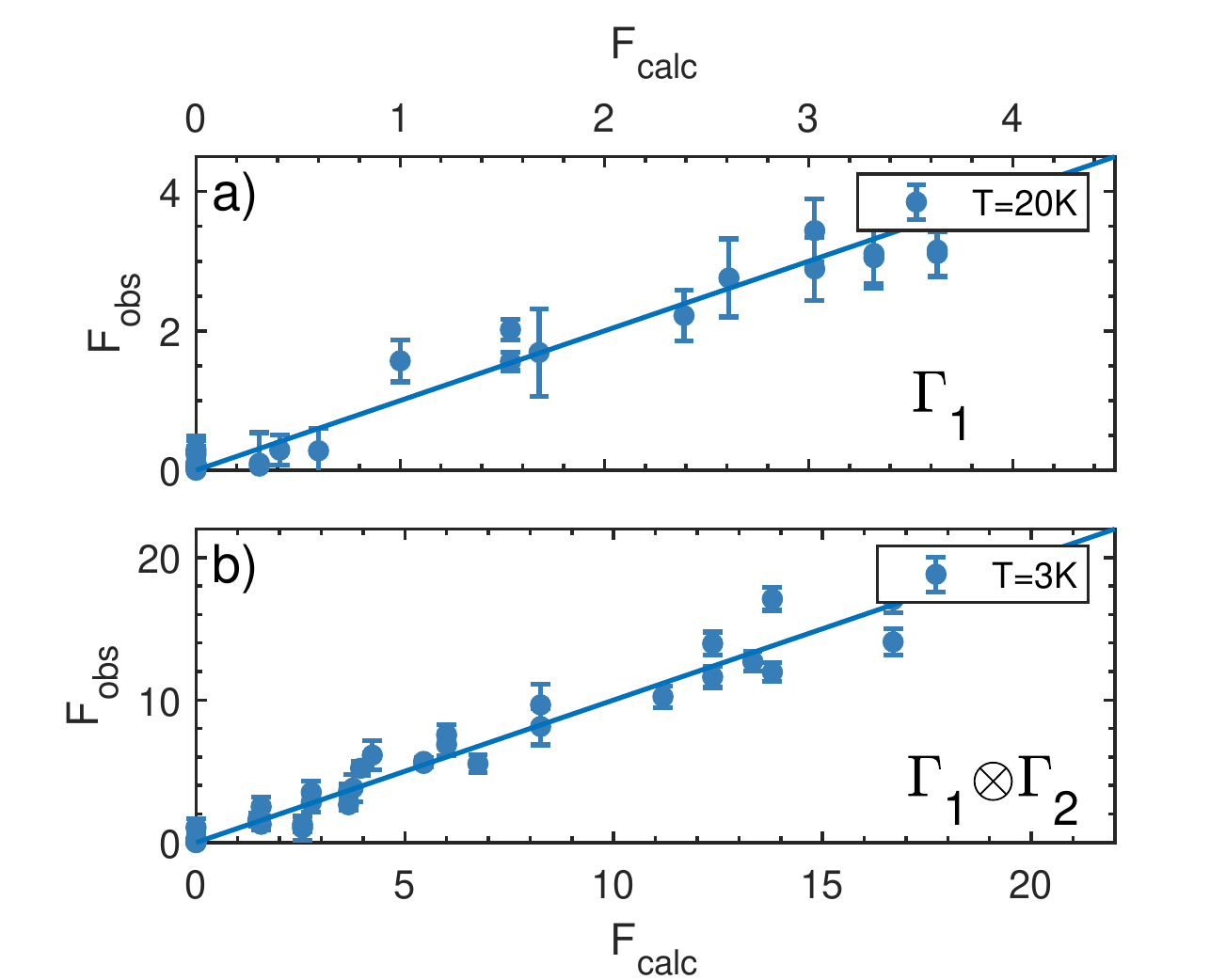}
  \caption{\label{fig:FOBS_FCALC}Both plots a) and b) display the observed structure factors against the calculated ones for the magnetic models $\Gamma_1$ and $\Gamma_1\otimes\Gamma_2$ at $T=\SI{20}{\kelvin}$ and $T=\SI{3}{\kelvin}$ respectively.}
  \label{disp}
 \end{figure}

The collection of magnetic reflections in zero-field for both phases IC1 and IC2 was carried out on the 6T2 diffractometer by using the same setup as for the temperature and magnetic-field dependent study of the incommensurability (see section \ref{sec:kvector}). Due to the instrumental setup the number of accessible reflections in $Q$ space was limited, wherefore only 25 magnetic reflections were recorded in the IC1 phase at $T=\SI{20}{\kelvin}$ and 39 magnetic reflections in the IC2 phase at $T=\SI{3}{\kelvin}$. The refinement and testing of different magnetic models was done with the software package \textsc{Fullprof}\cite{Carvajal1993}. All resulting reliability values for the refinements of respective models in both phases are displayed in table \ref{tab:reliability_values} and the observed versus calculated structure factors plots for the best resulting refinements are displayed in Fig. \ref{fig:FOBS_FCALC}.
\begin{table}
\caption{\label{tab:reliability_values} The reliability values of refinements assuming different magnetic models at $T=\SI{20}{\kelvin}$ and $T=\SI{3}{\kelvin}$ respectively are displayed in this table. The refinement was done by using the \textsc{Fullprof} suite\cite{Carvajal1993}; measured intensities were corrected for absorption effects. }
\begin{ruledtabular}
\begin{tabular}{ccccccc}

&$\SI{20}{\kelvin}$&&&$\SI{3}{\kelvin}$&&\\
&$\Gamma_1$&$\Gamma_2$&&$\Gamma_1$&$\Gamma_2$&$\Gamma_1\otimes\Gamma_2$\\\hline
$R_{F^2}$&16.2&92.9&&25.7&26.6&14.8\\
$R_{wF^2}$&20.2&87.6&&24.1&28.7&14.6\\
$R_{F}$&19.7&74.1&&19.8&17.5&13.2\\
$\chi^2$&1.43&26.7&&7.01&9.95&2.56\\
\end{tabular}
\end{ruledtabular}
\end{table}

The best refinement result for the IC1 phase was unambiguously achieved by assuming a magnetic model that is compatible with the single irreducible representation $\Gamma_1$. Since for this phase, the susceptibility data indicates  a magnetic arrangement exclusively lying within the $ac$ plane and in order to reduce the number of parameters, the $v$ component was fixed to zero. With this assumption, the magnetic moment on site 1 was refined to be $\vec{m}=\left(2.62(14)\:0\:2.26(18)\right)\mu_\text{B}$\: and thus the magnetic arrangement of the IC1 phase is described by a SDW that is confined to the $ac$ plane with an easy axis $\vec{e_{ac}}$ that forms an angle of about $\SI{41}{\degree}$\: with the $a$ axis (or $\SI{9}{\degree}$ with propagation vector). The alignment of moments almost bisecting the $a$ and $c$ directions nicely agrees with the anomalies appearing in the susceptibility for magnetic fields along $a$ and $c$.

\begin{figure}
 \includegraphics[width=\columnwidth]{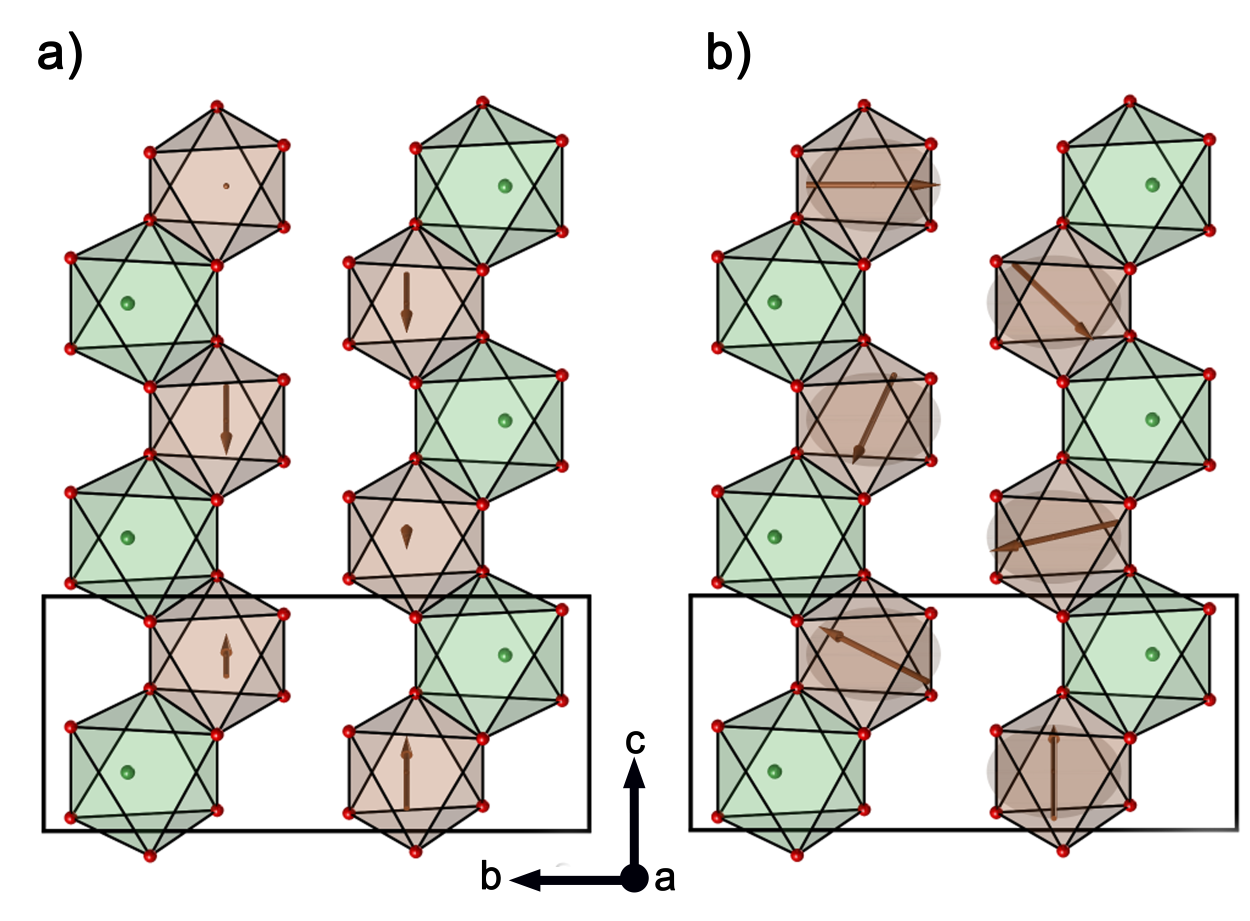}
  \caption{\label{fig:magnetic_structure}The panels a) and b) visualize the refinement results for the magnetic structure in the magnetic phases IC1 and IC2 respectively. For the sake of simplicity tungsten ions are not shown.}
  \label{disp}
 \end{figure}

Below $T_{\text{N}2}$ an additional $b$ component develops and the best refinement result was obtained by combining $\Gamma_1$ and $\Gamma_2$. The $a$ and $c$ components are described by $\Gamma_1$, whereas $\Gamma_2$ describes the $b$ component. Moreover, it turned out that choosing $u$, $w$ to be real and $v$ to be imaginary (corresponding to a $\pi/2$ phase shift) yield the lowest reliability values for the refinement, and thus the magnetic model following $\Gamma_1\otimes\Gamma_2$ describes an elliptical spin spiral arrangement for the IC2 phase with moments that are rotating within the plane spanned by $\vec{e_{ac}}$ and $\vec{b}$.
It has to be noted that the combination of two irreducible representations does not violate Landau theory, considering two consecutive second-order transitions, which is the typical scenario in many type-II multiferroics. The refinement yields $\vec{m}=\left(5.05(19)\:i4.21(24)\:2.93(27)\right)\mu_\text{B}$\: and the length of the major and minor principal axes of the elliptical spiral amounts $m_\text{max}=\sqrt{m_a^2+m_c^2}=5.84(21)\text{ }\mu_\text{B}$ and $m_\text{min}=m_b=4.21(24)\text{ }\mu_\text{B}$\: respectively and hence depict a moderate deformation with respect to a circular envelope. 
The fact that the larger axis of this ellipse is even larger than the full spin moment of Fe$^{3+}$ indicates some anharmonicity.
The $a,c$ principal axis of the spiral is now almost parallel to the propagation vector ($\SI{2}{\degree}$).
The averaged magnetic moment of the elliptical spiral amounts to $5.09(11)\text{ }\mu_\text{B}$ in perfect agreement with the Fe$^{3+}$ spin moment of S=5/2.
Both magnetic structures are visualized in Fig. \ref{fig:magnetic_structure}. Compared to its sister compound NaFe(WO$_4$)$_2$ the order moment is 
much larger due to the more threedimensional arrangement of magnetic coupling.
Also disorder seems not to play a crucial role in this material, see also the Appendix.
\life does not exhibit a lower commensurate phase as \mnwokomma{,} thus making its phase diagram simple and comparable to the zero-field behavior of multiferroric \nafegeo and \nh  \cite{Kim2012,Ding2018,Ackermann2013,Velamazan2015,Tian2016}.

%

  \subsection{\label{sec:neutron_polarization_analysis}Neutron polarization analysis}
Neutron polarization analysis can be used to separate the magnetic components and to sense the chirality of the magnetic structure. The respective experiment on \life was executed at the cold three axes spectrometer KOMPASS, located at the MLZ. The sample was mounted within the scattering plane $\left(1\:0\:0\right)/\left(0\:0\:1\right)$ and a Helmholtz coil setup was deployed for defining the guide field direction at the sample position. For a longitudinal polarization analysis the common right handed coordinate system was defined, for which $x$ is parallel to $\vec{Q}$, $y$ perpendicular to $x$ but within the scattering plane and $z$ perpendicular to $x$ and $y$. Thus, the crystallographic $b$ direction is aligned parallel to $z$ direction. The small sample volume demanded a high neutron flux. Therefore, the horizontal collimation in front of the secondary spin-analyzing cavity was removed. As the efficiency of the V-shaped cavity depends on the incoming beam divergence, the flipping ratio (FR) was reduced significantly to FR=8-11.
\begin{figure}
 \includegraphics[width=\columnwidth]{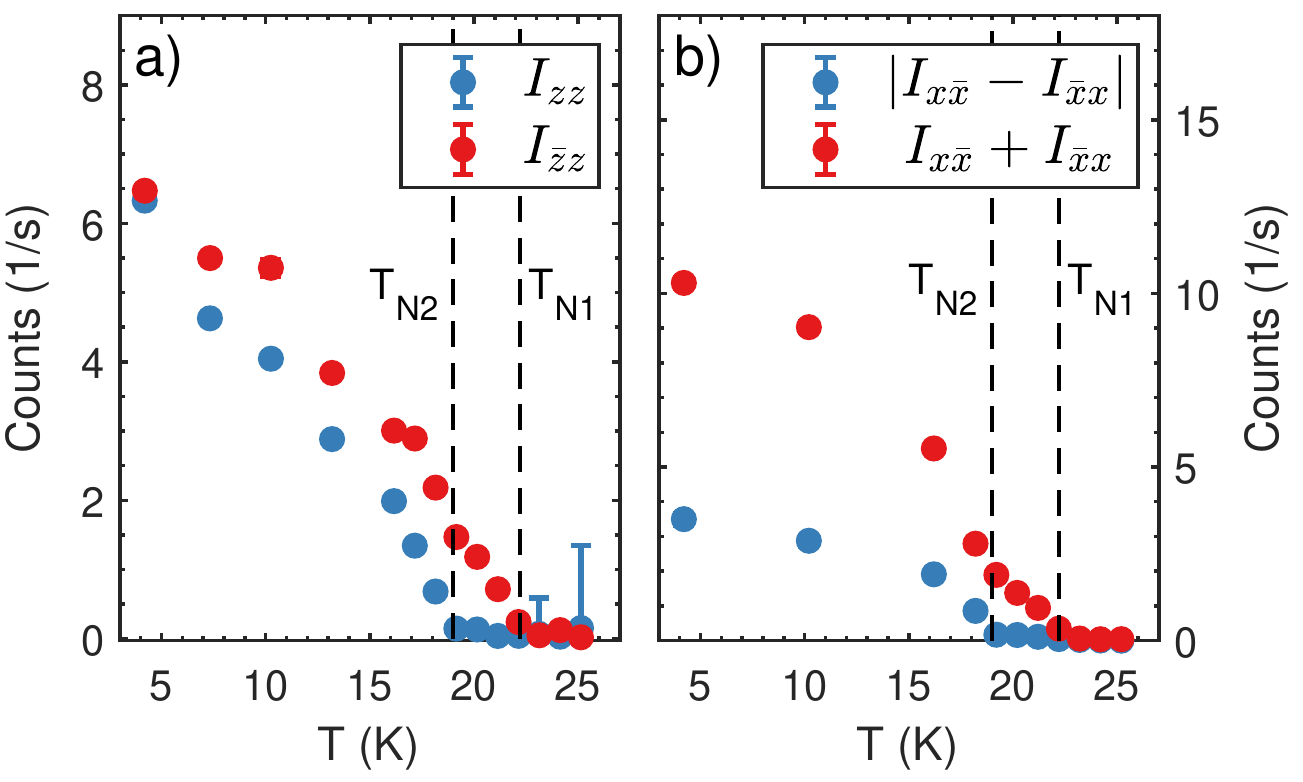}
  \caption{\label{fig:SF_NSF}The displayed intensities in a) correspond to the recorded NSF and SF channels $I_{zz}$ and $I_{\bar{z}z}$ with $\vec{P}||z$ and $\vec{Q}=\left(-1.09\:0\:0.30\right)$. In b) the sum and the difference of both SF channels with $\vec{P}||x$ are plotted.}
  \label{disp}
 \end{figure}

In the first part of the experiment, the neutron beam polarization $\vec{P}$ was aligned parallel to $z$  and a single spin-flipper was placed in front of the sample thus allowing to measure the SF channel $I_{\bar{z}z}$ and the NSF channel  $I_{zz}$. The intensities for the respective SF and NSF channels were recorded for the magnetic reflection $\vec{Q}=\left(-1.09\:0\:0.30\right)$ and their temperature dependence is displayed in Fig. \ref{fig:SF_NSF}a). Due to the chosen position in $\vec{Q}$ space, no nuclear Bragg signal contributes to the scattered intensity. In general, only the magnetization $\vec{M_\perp}(\vec{Q})$ perpendicular to $\vec{Q}$ contributes to the scattering process. Moreover, a neutron spin-flip requires a component of the magnetization perpendicular to the neutron polarization, whereas non-spin flip processes occur, when the neutron polarization is collinear to the magnetization. Here, it entails that the respective SF and NSF channels are described by $I_{zz}=M_{b}M_{b}^*$ and $I_{\bar{z}z}=\sin^2(\alpha)M_{ac}M_{ac}^*$ with $M_{b}$ and $M_{ac}$ being the complex Fourier components of the magnetization along $b$ direction and within the $ac$ plane, respectively, and $\sin^2(\alpha)$ being the geometry factor. From Fig. \ref{fig:SF_NSF}a) it can be clearly stated that in the IC1 phase a magnetic component is solely evolving within the $ac$ plane. Below $T_{\text{N}2}$ an additional $b$ component develops and its magnitude approaches a similar size with respect to the $ac$ component, when lowering the temperature. The respective temperature dependent ratio of both components confirms the refinement results for both phases discussed above.

In the second part of the experiment, the neutron beam polarization was set parallel to $x$ (parallel to $\vec{Q}$)
and a second spin-flipper was placed behind the sample. With this configuration both SF channels  $I_{x\bar{x}}=\vec{M_\perp}\vec{M_\perp^*}-i\left(\vec{M_\perp}\times \vec{M_\perp^*}\right)_x$ and  $I_{\bar{x}x}=\vec{M_\perp}\vec{M_\perp^*}+i\left(\vec{M_\perp}\times \vec{M_\perp^*}\right)_x$ can be measured giving access to the chiral component $\pm i\left(\vec{M_\perp}\times \vec{M_\perp^*}\right)_x$, for which the sign depends on the vector chirality of the spiral. Fig. \ref{fig:SF_NSF}b) displays the sum and the difference of both SF channels hence presenting the temperature dependence of the overall magnetic scattering contribution $2\vec{M_\perp}\vec{M_\perp^*}$ and that of the chiral signal $-i2\left(\vec{M_\perp}\times \vec{M_\perp^*}\right)_x$, respectively.
It is clearly visible that the magnetic signal develops below $T_{\text{N}1}$, whereas the chiral structure appears only when passing the lower transition at $T_{\text{N}2}$. This is in accordance with the discussed refinement results, which proposed a SDW and a spiral spin arrangement below $T_{\text{N}1}$ and $T_{\text{N}2}$, respectively. Astonishingly, the finite value of $|I_{x\bar{x}}-I_{\bar{x}x}|$ develops without an applied external electric field. The sample crystal thus exhibits an intrinsic preferred vector chirality, as it has also been reported for other type-II multiferroics \cite{Finger2010}.

\subsection{\label{sec:discussion}Discussion of magnetic interaction and multiferroic coupling}

The transition temperatures for both observed phases IC1 and IC2 are significantly larger than those reported in \nafe and \mnwokomma{.}
The low transition temperatures in \nafe can be attributed to the reduced interlayer coupling arising from the separation through nonmagnetic Na layers. However, the lower transition temperatures in MnWO$_4$ disagree with the higher density of magnetic Mn$^{2+}$ ions compared to that of Fe$^{3+}$ ions in \lifekomma{,} which possess the same $3d^5$ configuration. The lower transition temperatures in MnWO$_4$ result from a higher degree of frustration \cite{Ye2011}.

The magnetic coupling between Fe$^{3+}$ ions in \life  is mediated through super-super-exchange interaction along Fe-O-O-Fe and Fe-O-W-O-Fe paths, which are marked by $J_1$, $J_2$ and $J_{3a}$/$J_{3b}$ in Fig. \ref{fig:structure} (d) and (e). The exchange coupling $J_{3}$ is split into $J_{3a}$ and $J_{3b}$ due to the monoclinic angle $\beta=\SI{90.574(2)}{\degree}$. The refinement based on single-crystal X-ray data at room temperature yields the inter- and intralayer distances between nearest-neighbor Fe ions of about $\SI{4.4970(10)}{\angstrom}$, $\SI{4.9026(4)}{\angstrom}$ and $\SI{5.5773(5)}{\angstrom}$/$\SI{5.6180(5)}{\angstrom}$   for the pairs described by $J_1$, $J_2$ and $J_{3a}$/$J_{3b}$, respectively. The distances of the pairs $J_1$ and $J_2$ are comparable so that a single zigzag chain is not the main magnetic unit. Instead the Fe$^{3+}$ of two neighboring zigzag chains together form a magnetic chain with nearest-neighbor interaction $J_1$ and next-nearest neighbor interaction $J_2$, which is a
classical configuration of frustration, see Fig. 1.

DFT calculations find $J_1$ to be ferromagnetic and $J_2$ and $J_{3a}$/$J_{3b}$ to be antiferromagnetic yielding magnetic frustration \cite{Liu2017}. The ferromagnetic coupling for $J_1$ can be verified by calculating the magnetic
exchange energy in the incommensurate phase. For the SDW and the spiral phases the classical exchange energy  per Fe resulting
from $J_1$ and $J_2$ amounts to:

\begin{equation}
E_{\text{exch}}=-2J_1\text{cos}(\pi q_c)-2J_2\text{cos}(2\pi q_c)
\end{equation}

In this phenomenological simple approach only a ferromagnetic (positive) $J_1$ with $J_2/J_1\approx-0.4$ agrees with the observed incommensurability along $c$ of $\sim$0.3 in phases IC1 and IC2 (see section \ref{sec:kvector}), which corresponds to a long wave length of about six Fe moments. A ferromagnetic coupling
between Fe$^{3+}$ moments is rather unusual in view of the half-filled $3d$ shell, but arises from cancelation of different super-exchange paths. The ferromagnetic
nearest-neighbor coupling is however much smaller in size than usual antiferromagnetic couplings so that the more distant interaction parameters
play an important role in \life . Furthermore, \life is not a highly one-dimensional system as the coupling between the double-zigzag chains is sizeable.

One may extend the exchange energy to $J_{3a}$/$J_{3b}$ ignoring the small difference arising from the monoclinic angle by adding a term:

\begin{equation}
E_{\text{exch}-J_3}=-2J_3[\text{cos}(\pi q_a+\pi q_c)+\text{cos}(\pi q_a-\pi q_c)].
\end{equation}

A finite antiferromagnetic $J_3$ stabilizes antiferromagnetic stacking along the $a$ direction corresponding to $q_a$=1 and reduces the $q_c$ value of
minimum exchange energy. For the parameters calculated within density-functional theory \cite{Liu2017} $J_1$=2.28\,meV, $J_2$=-2.44\,meV and $J_3$=-1.13\,meV
the minimum occurs at $q_c$=0.347 close to the experimental value of 0.30 while the minimum appears at 0.425 without taking $J_3$ into account.
In order to explain the incommensurate modulation along $a$ one needs to take interaction at even farer distances into account. There are 4 neighbors
at $\SI{7.33}{\angstrom}$ distance, but this shell corresponds to the $C$ centering vector and thus enforces $q_a$=1 for an antiferromagnetic parameter.
The next 6 Fe-Fe shells are at distances between 8 and $\SI{9}{\angstrom}$ with in total 16 bonds. Therefore, a reasonable estimate of the impact
of these interaction parameters cannot be made.

In contrast to \nafe \cite{Holbein2016} no second intrachain spiral with opposite handedness exists as the upper and lower rows of a zigzag chain are alternatingly occupied by magnetic Fe$^{3+}$ and nonmagnetic Li$^{1+}$. Furthermore spirals of neighboring chains possess the same handedness, wherefore the effect of inverse DMI is not canceled in \lifekomma{.} It is the combination of the two incommensurate magnetic modes that breaks the inversion symmetry in \lifekomma{.}
For the inverse DMI the ferroelectric polarization is given by
\begin{equation}
 \vec{P}\propto\vec{r_{ij}}\times\left(\vec{S_i}\times\vec{S_j}\right)
\end{equation}
with $\vec{S_i}$ and $\vec{S_j}$ being two neighboring spins and $\vec{r_{ij}}$ being the connecting vector of them \cite{Mostovoy2006,Kimura2007}.
With this formalism we can determine the direction of the ferroelectric polarization.
With the propagation vector $\vec{k_\text{inc}}=\left(0.91\:0\:0.30\right)$ being perpendicular to $b$ and with the spiral structure arising from $e_{ac}$ and $b$ components it is obvious that the four  Fe moments coupled through $J_1$ and $J_2$ yield a ferroelectric polarization along the $b$ direction. Taking into account all other pairs described by adding translations $\pm(n\vec{a}+m\vec{b}+l\vec{c})$ also shows that any polarization contribution perpendicular to $b$ cancels out.
The same conclusion can be deduced from the symmetry of the magnetic spiral combining two representations so that the $c$ glide mirror plane perpendicular is broken.

 \section{\label{sec:level1} Conclusions}

We present a comprehensive single-crystal investigation of the magnetization, the specific heat under magnetic field, and of the nuclear and magnetic structure of the newly discovered multiferroic material \life \cite{Liu2017}. Temperature dependent susceptibility and specific heat measurements reveal the magnetic anisotropy.
The system undergoes two magnetic transitions at $T_{\text{N}1}\approx\SI{22.2}{\kelvin}$ and $T_{\text{N}2}\approx\SI{19}{\kelvin}$. 
With single-crystal neutron diffraction we were able to observe long-range incommensurate magnetic ordering not only in the multiferroic but also in the intermediate phase, which so far has been proposed to exhibit only short-range ordering \cite{Liu2017}. The incommensurability in both magnetic phases shows only a slight temperature dependence, whereas the magnetic field up to $\SI{6}{\tesla}$ does not change the magnetic propagation vector at all. It was possible to determine and refine the magnetic structure in both phases IC1 and IC2, revealing a SDW with an easy axis lying within the $ac$ plane in phase IC1 and an elliptical spiral with an additional $b$ component in phase IC2. Both refinements, in particular the chiral nature of the IC2 magnetic structure, are confirmed by neutron polarization analysis in respective phases. The refined chiral structure of the IC2 phase is compatible with the proposed ferroelectric polarization along $b$ direction arising from the inverse DMI \cite{Liu2017,Mostovoy2006,Kimura2007} and it was observed that in IC2 an unbalanced multiferroic domain distribution appears even in the absence of an applied external electric field. No transitions to commensurate phases were observed, rendering the phase diagram simple
and \life \ a well suited material to study intrinsic multiferroic properties such as domain dynamics. Due to the particular arrangement of magnetic Fe ions in the zigzag octahedron chains, \life can be moreover  considered as a realization of a $J_1$,$J_2$ magnetic chain, although the interchain interaction is
sizeable and results in a fully ordered structure at low temperature.
Both, the simple phase diagram and the threedimensional coupling of magnetic zigzag chains  encourage further research on multiferroic domain dynamics and magnetic interactions in \lifekomma{.}


\section{\label{sec:level1}Acknowledgements}


This work was funded by the Deutsche Forschungsgemeinschaft (DFG,
German Research Foundation) - Project number 277146847 - CRC 1238, projects A02, B01 and B04 and by
the Bundesministerium f\"ur Bildung und
Forschung - Project number 05K19PK1.

\appendix
\section{\label{sec:appendix}}

\begin{figure}
 \includegraphics[width=\columnwidth]{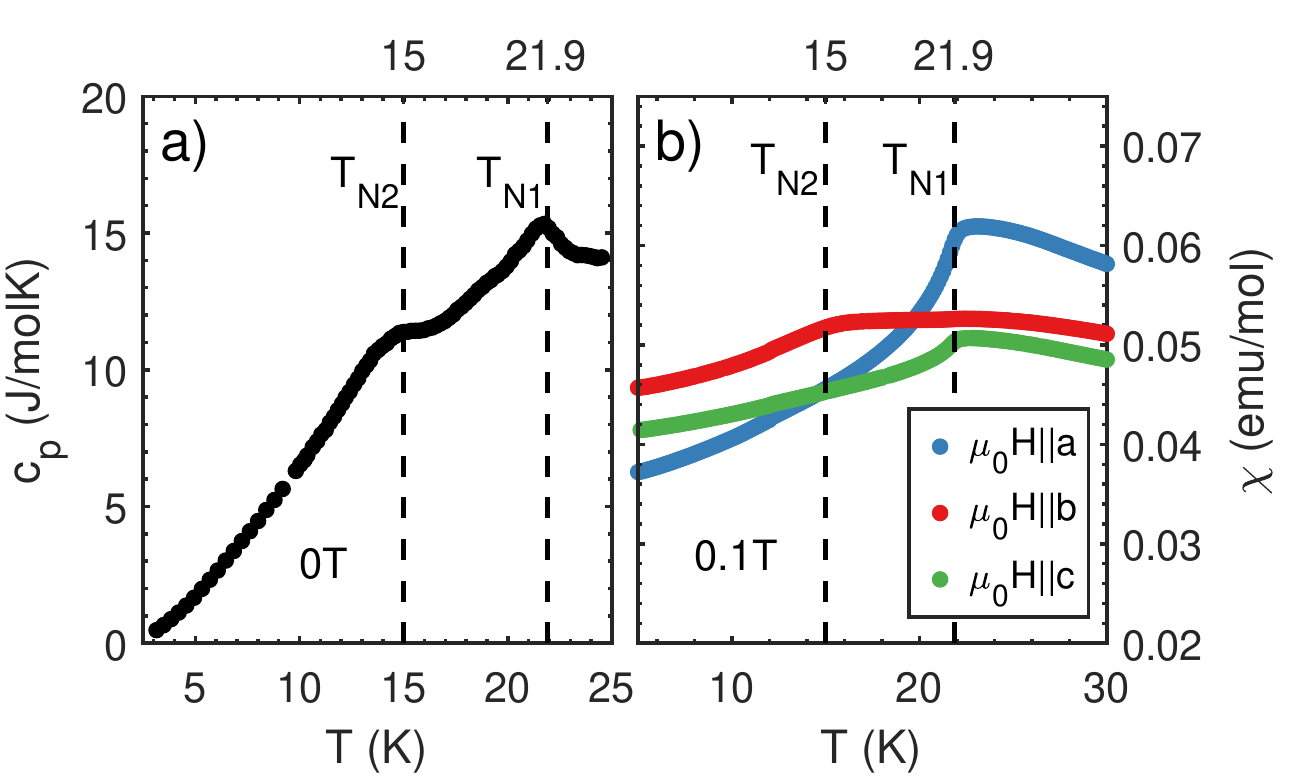}
  \caption{\label{fig:SII_cp_susceptibility}The specific heat and susceptibility measurements on sample SII reveal reduced transition temperatures $T_{\text{N}1}$ and $T_{\text{N}2}$ compared to sample SI, which are related to Li/Fe site disorder (see text).}
  \label{disp}
 \end{figure}

Characterization measurements on several \life samples revealed two different sample types that exhibit different transition temperatures. For some samples the specific heat as well as the susceptibility measurements display a slightly lowered value for $T_{\text{N}1}$, whereas the $T_{\text{N}2}$ value was diminished significantly to $\approx\SI{15}{\kelvin}$. Exemplary measurements are shown for sample SII in Fig. \ref{fig:SII_cp_susceptibility}. Sample SII was investigated furthermore by neutron diffraction at the four circle instrument HEiDi \cite{Meven2015}, which is located at the MLZ and jointly operated by RWTH Aachen University and Forschungszentrum J\"ulich GmbH within JARA-FIT collaboration. A combined set of nuclear reflections was recorded with $\lambda=\SI{0.795}{\angstrom}$ and $\lambda=\SI{1.171}{\angstrom}$ utilizing a germanium monochromator (Ge(422) and Ge(311) respectively). The refinement of the nuclear structure yields similar results for the atomic positions (see table  \ref{tab:table_nuc_refinement_SII}) compared to the structural refinement of sample SI (see table  \ref{tab:table_nuc_refinement}).
The onset of magnetic order does not yield a significant change in the crystal structure.
However, the structural refinement of sample SII discloses a mis-occupation of the Li-site by Fe ions of about $\approx\SI{6.7(5)}{\percent}$ at $\SI{24}{\kelvin}$ and $\approx\SI{6.2(4)}{\percent}$ at $\SI{2.5}{\kelvin}$, which provokes a strong impact on magnetic interactions and hence on the respective magnetic ordering and transition temperatures. In contrast, the X-ray experiment on sample SI yields an insignificant misoccupation of only about 0.4(5) percent. Thus, the discrepancy concerning the observed magnetic transition temperatures for both sample types and with respect to the reported phase diagram \citep{Liu2017} of multiferroic \life arises from the alteration of the magnetic zigzag chains through a misoccupation by Fe on the Li site. However, in both phases of SII, a magnetic reflection indexed by a similar propagation vector as for SI was detectable.

\begin{table}
\caption{\label{tab:table_nuc_refinement_SII}This table shows nuclear refinements based on neutron diffraction data from sample SII for two temperatures. The Li-site was refined, while assuming a mis-occupation of it by Fe ions ($\approx\SI{6.7(5)}{\percent}$ at $\SI{24}{\kelvin}$ and $\approx\SI{6.2(4)}{\percent}$ at $\SI{2.5}{\kelvin}$). The refinement was done on structure factors and with the software \textsc{Jana2006} \cite{Petricek2014}.  The integrated intensities were numerically corrected for absorption and an extinction
correction was applied during the refinements.}
\begin{ruledtabular}
\begin{tabular}{m{0.9cm}ccccc}

 &&x  &y  & z &U$_\text{iso}$\\ \hline
$\SI{24}{\kelvin}$&W&0.24755(19)&0.09125(13)&0.2470(4)&0.0029(4)\\
 &Fe&0.0  &0.33480(9)  &0.25  &0.0024(3)\\
      &Li&0.5  &0.3495(10)   &0.25  &0.016(4)\\
      &O1&0.36287(13)  &0.05852(10)  &0.9250(3)  &0.0043(4)\\
      &O2&0.38014(12)  &0.18270(11)  &0.4102(3)  &0.0045(4)\\
      &O3&0.35553(13)  &0.54899(10)  &0.9431(3)  &0.0039(4)\\
      &O4&0.37747(12)  &0.69384(10)  &0.3926(3)  &0.0041(4)\\
 &\multicolumn{5}{m{7cm}}{\centering Recorded reflections: 578, Independent: 290 R(obs)=1.99, wR(obs)=2.19, R(all)=2.24, wR(all)=2.22
}\\
  &&&&&\\
$\SI{2.5}{\kelvin}$&W&0.24754(15)&0.09120(11)&0.2472(3)&0.0029(3)\\
 &Fe&0.0  &0.33469(8)  &0.25  &0.0022(3)\\
      &Li&0.5  &0.3493(9)   &0.25  &0.019(3)\\
      &O1&0.36294(11)  &0.05853(9)  &0.9252(2)  &0.0042(3)\\
      &O2&0.38028(10)  &0.18278(9)  &0.4102(2)  &0.0045(3)\\
      &O3&0.35572(11)  &0.54913(9)  &0.9429(2)  &0.0035(3)\\
      &O4&0.37738(10)  &0.69370(9)  &0.3924(2)  &0.0044(3)\\
 &\multicolumn{5}{m{7cm}}{\centering Recorded reflections: 1259, Independent: 730 R(obs)=2.42, wR(obs)=2.37, R(all)=2.42, wR(all)=2.37
}\\
\end{tabular}
\end{ruledtabular}
\end{table}

\newpage
\nocite{apsrev41Control}
\bibliographystyle{apsrev4-1}

\end{document}